\documentclass{elsart3}

\usepackage{amssymb}
\usepackage{graphicx}

\begin{document}

\begin{frontmatter}



\title{Vertex reconstruction algorithms in the PHOBOS experiment at RHIC}


\thanks[collaboration]{
\footnotesize
%
%
%
B.Alver$^4$,
B.B.Back$^1$,
M.D.Baker$^2$,
M.Ballintijn$^4$,
D.S.Barton$^2$,
R.R.Betts$^6$,
A.A.Bickley$^7$,
R.Bindel$^7$,
A.Budzanowski$^3$,
W.Busza$^4$,
A.Carroll$^2$,
Z.Chai$^2$,
V.Chetluru$^6$,
M.P.Decowski$^4$,
E.Garc\'{\i}a$^6$,
T.Gburek$^3$,
N.George$^2$,
K.Gulbrandsen$^4$,
S.Gushue$^2$,
C.Halliwell$^6$,
J.Hamblen$^8$,
G.A.Heintzelman$^2$,
C.Henderson$^4$,
I.Harnarine$^6$,
D.J.Hofman$^6$,
R.S.Hollis$^6$,
R.Ho\l y\'{n}ski$^3$,
B.Holzman$^2$,
A.Iordanova$^6$,
E.Johnson$^8$,
J.L.Kane$^4$,
N.Khan$^8$,
W.Kucewicz$^6$,
P.Kulinich$^4$,
C.M.Kuo$^5$,
W.Li$^4$,
W.T.Lin$^5$,
C.Loizides$^4$,
S.Manly$^8$,
A.C.Mignerey$^7$,
R.Nouicer$^{2,6}$,
A.Olszewski$^3$,
R.Pak$^2$,
I.C.Park$^8$,
C.Reed$^4$,
L.P.Remsberg$^2$,
M.Reuter$^6$,
E.Richardson$^7$,
C.Roland$^4$,
G.Roland$^4$,
L.Rosenberg$^4$,
J.Sagerer$^6$,
P.Sarin$^4$,
P.Sawicki$^3$,
I.Sedykh$^2$,
W.Skulski$^8$,
C.E.Smith$^6$,
M.A.Stankiewicz$^2$,
P.Steinberg$^2$,
G.S.F.Stephans$^4$,
A.Sukhanov$^2$,
A.Szostak$^2$,
J.-L.Tang$^5$,
M.B.Tonjes$^7$,
A.Trzupek$^3$,
C.Vale$^4$,
G.J.van~Nieuwenhuizen$^4$,
S.S.Vaurynovich$^4$,
R.Verdier$^4$,
G.I.Veres$^4$,
E.Wenger$^4$,
D.Willhelm$^2$,
F.L.H.Wolfs$^8$,
B.Wosiek$^3$,
K.Wo\'{z}niak$^3$,
A.H.Wuosmaa$^1$,
S.Wyngaardt$^2$,
B.Wys\l ouch$^4$\\
$^1$~Argonne National Laboratory, Argonne, IL, USA\\
$^2$~Brookhaven National Laboratory, Upton, NY, USA\\
$^3$~Institute of Nuclear Physics PAN, Krak\'{o}w, Poland\\
$^4$~Massachusetts Inst.~of Technology, Cambridge, MA, USA\\
$^5$~National Central University, Chung-Li, Taiwan\\
$^6$~University of Illinois at Chicago, Chicago, IL, USA\\
$^7$~University of Maryland, College Park, MD, USA\\
$^8$~University of Rochester, Rochester, NY, USA
%
}

\author{Krzysztof Wo\'{z}niak for the PHOBOS Collaboration\thanksref{collaboration} }
\address{Institute of Nuclear Physics PAN, Krak\'{o}w, Poland }

\begin{abstract}
{
The PHOBOS experiment at the Relativistic Heavy Ion Collider (RHIC) at
Brookhaven National Laboratory is studying interactions of heavy nuclei
at the largest energies available in the laboratory. 
The high multiplicity of particles created in heavy ion collisions makes
precise vertex reconstruction possible using information from a
spectrometer and a specialized vertex detector with relatively small
acceptances. For lower multiplicity events, a large acceptance, single
layer multiplicity detector is used and special algorithms are developed to
reconstruct the vertex, resulting in high efficiency at the expense of
poorer resolution. The algorithms used in the PHOBOS experiment and their
performance are presented.
}
\end{abstract}

\end{frontmatter}

%
\begin{figure}[hbt]
\begin{center}
\includegraphics[width=7.5cm]{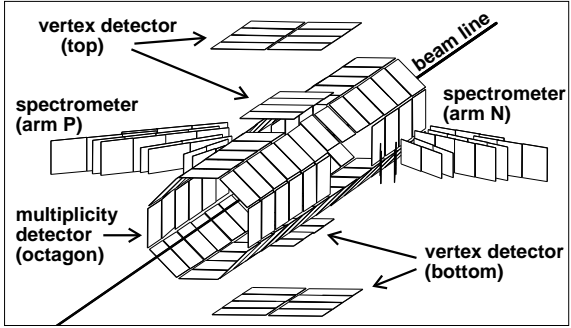}
\end{center}
\caption{ \label{Fig1}
The central part of the PHOBOS detector \cite{detector}.
Only the silicon sensors of the octagonal multiplicity detector, 
the vertex detector and the eight first planes of the two spectrometer arms are shown. 
In the PHOBOS coordinates system mentioned in the text, the $z$ axis
coincides with the beam line, the $y$ axis points upward and the 
$x$ axis is horizontal in the direction of spectrometer arm P.
}
\end{figure}
\section{Introduction}
The  PHOBOS experiment is studying 
Au+Au, Cu+Cu, d+Au and p+p collisions
at the nucleon-nucleon center of mass energies from
$\sqrt{s_{NN}} = 19.6$ GeV to $\sqrt{s_{NN}} = 200$ GeV
(and $\sqrt{s_{NN}} = 410$ GeV for protons).
For an overview of results see \cite{whitepaper}, for 
the description of the detector see \cite{detector,sensors}.
General properties of these collisions are measured
using a large acceptance multiplicity detector covering almost the full solid angle.
About 50\% of the produced particles are registered in the central, octagonal part
of multiplicity detector (Fig.~1), whereas those emitted at smaller angles 
with respect to the beam line are detected in the multiplicity ring detectors
placed a~few meters from the nominal interaction point 
(not shown in Fig.~1).
Detailed information on about 1\% of the particles 
is obtained from a two-arm magnetic spectrometer.
The position of the primary interaction is precisely reconstructed using a
specialized vertex detector, which registers about 5\% of the primary particles.
At RHIC, the beams of ions collide at zero degree, thus 
the primary vertices are distributed in a wide $z$~range along the beam line 
(up to $\pm$2~m from the nominal interaction point) and only
about $\pm$0.1~cm in the perpendicular direction.
The relatively small acceptance of the vertex detector is 
sufficient for vertex reconstruction 
for events with large multiplicities,
but does not allow to extend the analysis to more peripheral Au+Au collisions 
or to the interactions of lighter systems, especially p+p.
Therefore, in order to improve the vertex determination efficiency and accuracy,
several algorithms, presented in this paper,
are needed using complementary information from 
different detector subsystems.
\section{Vertex reconstruction algorithms}
The scintillator counters of the trigger system (not shown in Fig.~1)
deliver the first information on the position of the primary vertex.
Two larger detectors, paddle counters, are placed symmetrically at 
$z$~=~$\pm$3.2~m from the nominal interaction point, 
two smaller T0 counters are mounted 
around the beam pipe at $z$~=~$\pm$5.3~m   
for most measurements (in p+p collisions they are moved closer, 
to $z$~=~-1.3~m and $z$~=~+2.5~m).
The coincidence of the signals from these detector pairs on different $z$-sides 
is used for triggering,
but as the arrival times of produced particles are measured,
the vertex $z$~position can be derived from the registered time difference.
The error in the vertex position determination (about 5~cm) 
is small enough for an on-line vertex range selection,
but it is not sufficient for data analyzes.

The vertex position in 3 dimensions ($X_v, Y_v, Z_v$) 
can only be obtained from tracks reconstructed in the PHOBOS spectrometer. 
It consists of 16~layers of silicon pad and strip sensors places inside
the conventional 2~Tesla magnet. The design of the magnet ensures that 
the first 8~layers of spectrometer silicon sensors are in an
area where the field is very weak. This allows easier reconstruction of 
the first segment of the particle trajectory as a straight line, 
which is then normally joined with a curved 
part of the track found in the remaining layers. 
The acceptance for the straight line segments is larger than 
that for the fully reconstructed tracks and the information on the curved
part does not improve track extrapolation to the vertex, thus 
the straight lines are used in two algorithms of vertex reconstruction.

In the {\it spectrometer 3D} method, the
vertex position is defined as the point for which the sum of distances
from the extrapolated tracks is minimal. In an iterative procedure,
tracks too distant from the calculated vertex (eg.~secondary particles) 
are removed and the vertex is recalculated. The {\it spectrometer 2D+1D}
algorithm enters the points of closest approach of all track pairs  
into 2-dimensional ($x-z$)  and 1-dimensional ($y$)
histograms. The coordinates of the maxima in both histograms define the approximate 
vertex position. The final vertex position is determined by the mean in $x$, $y$ and $z$
of only those points compatible with the beam orbit and close enough to the approximate 
vertex. The sizes of the pads in the silicon
sensors used for vertex reconstruction are relatively large 
(1$\times$1 mm$^2$ and 0.427$\times$6 mm$^2$).
This, together with small acceptance of the spectrometer, is responsible
for the limited precision of the spectrometer methods.

 \begin{figure}[tbh]
 \includegraphics[width=7.5cm]{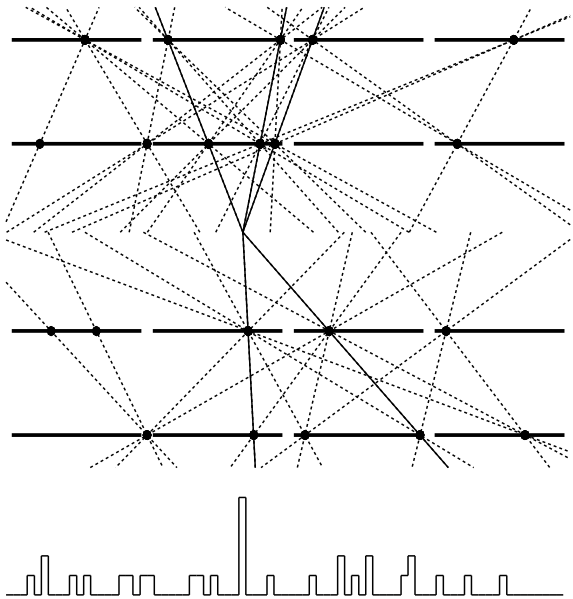}
 \caption{ \label{Fig2}
 Determination of the vertex position using all combination of hits
 registered in the vertex detector. Dashed lines represent false 
 combinations, continuous lines primary particles. The histogram
 of the number of tracks extrapolated to a fixed $y$ value 
 (at the bottom of the picture) is used to
 find the position of the vertex. }
 \end{figure}

The PHOBOS vertex detector consists of two parts placed above and below the beam pipe,
25~cm long in~$z$, each with 2 layers of silicon sensors
at about 5.7~cm and 11.8~cm from the beam. The sensors have strips 473 $\mu$m wide 
and 1.2 or 2.4~cm long, perpendicular to the beam direction. The strips in the more
distant outer layers are two times longer to cover the same azimuthal angle range.
The design of the vertex detector limits the vertex reconstruction  
to $y$ and $z$~positions. In the vertex search 
all combinations of hits in the two layers are treated as track candidates and their 
extrapolated $z$ position at a fixed $y$ is histogrammed (Fig.~2).  
$Z_v$ is defined by the position of the maximum of such a histogram.
This procedure is repeated for several values of fixed $y$.
The value of $y$ for which the maximum of the histogram is the highest 
is used as final $Y_v$. 
To accept the maximum at least three track candidates contributing to it are
required, since without this cut coincidences of two false track
candidates are selected in low multiplicity events more
frequently than the true vertex.

\begin{figure}[htb]
\includegraphics[width=7.5cm]{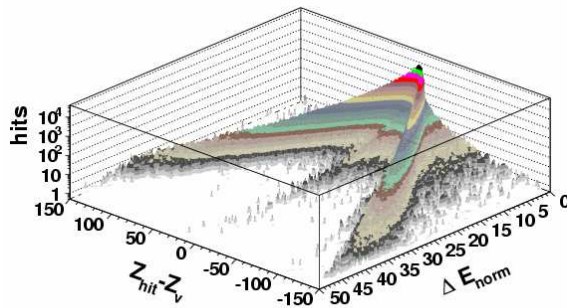}
\caption{ \label{Fig3}
The correlation of the difference between the position of the hit 
and the position of the primary vertex, 
$Z_{hit}-Z_{v}$, and the registered energy loss $\Delta E_{norm}$
in the PHOBOS octagonal multiplicity detector.
To correct for the different sensor thicknesses, the measured energy loss
is divided by the mean energy of a minimum ionizing particle traversing
the sensor perpendicularly.}
\end{figure}

In PHOBOS, the largest fraction of primary particles 
is registered in the octagonal multiplicity
detector consisting of 92 silicon sensors with 0.27$\times$0.86 cm$^2$ pads.
Their thickness varies from 0.305~mm to 0.338~mm.
The sensors are placed around the beam pipe at radii of about 4.5~cm
and cover the $z$~range from -55~cm to +55~cm.  
In this detector primary particles traverse only one silicon sensor and
do not leave a track directly pointing to the vertex. 
However, the length of their trajectory inside
the silicon (and thus the deposited ionization energy, $\Delta E$) 
depends on the angle at which they enter the sensor; for primary particles  
approximately the emission angle. 
In this way the measured energy loss is correlated with the distance of the hit from
the vertex (Fig.~3). Most of the hits belong to one of the two ridges,
symmetric with respect to $Z_{hit}-Z_{v}$~=~0. 
The hits left by particles with very low
momenta, which deposit much larger energy than the minimum ionizing particles,
are visible at $Z_{hit}-Z_{v}$ close to zero as a tail of large $\Delta E$ values.
Because of the positive/negative $Z_{hit}-Z_{v}$ ambiguity, 
the value of $\Delta E$ from a single hit can be used only to estimate 
the distance to the vertex or two alternative vertex positions.
To determine the $z$ vertex position slices with fixed $\Delta E$ values
are calculated from the 2-dimensional histogram shown in Fig.~3.
Several examples of these histograms are shown in Fig.~4.
Using such histograms the probability density function, 
$P(\Delta E,|Z_{hit}-Z_{v}|)$, for a particle that loses the energy
$\Delta E$ at the position $Z_{hit}$ to originate from 
the vertex at $Z_{v}$ is calculated.
 This function is non-zero in two $z$ ranges, symmetric with respect to $Z_{hit}$.
For most registered primary particles the width of a single range is 3-6~cm,
and useful information is provided by the hits at  
the distance $|Z_{hit}-Z_{v}|$ up to 100~cm. 
For a single hit there is no way to select the correct range, and one
needs an overlap of several ranges from more hits to find
the correct vertex position.
The $z$ position of the vertex is determined by maximizing the product
of the probabilities of all hits in the octagonal multiplicity detector.

\begin{figure}[htb]
\includegraphics[width=7.5cm]{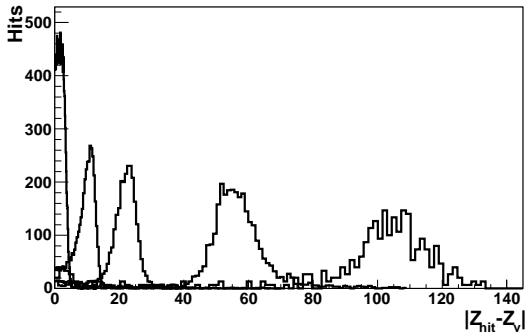}
\caption{ \label{Fig4}
Distributions of the
distances between the primary vertex and the hits for several different
values of the normalized energy loss $\Delta E_{norm}$.
The histograms (from left to right) are obtained for $\Delta E_{norm}$ equal
1.2, 3, 6, 15 and 30.}
\end{figure}

\section{Vertex reconstruction quality}

In the analysis of the vertex reconstruction quality we will focus first on
Au+Au collisions at the highest RHIC energy, $\sqrt{s_{_{NN}}}$ = 200 GeV,
using Monte Carlo simulations of the PHOBOS detector response based on GEANT 3.21
package \cite{geant}.

\begin{figure}[bt]
\includegraphics[width=7.5cm]{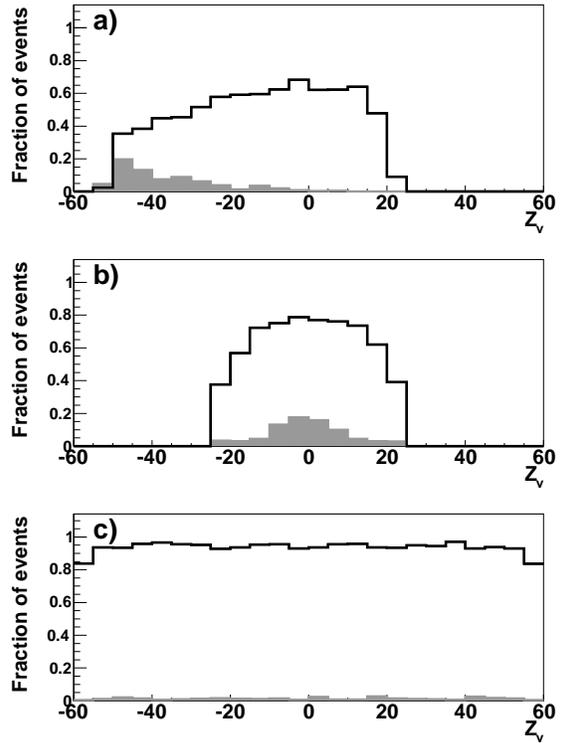}
\caption{ \label{Fig5}
The vertex reconstruction efficiency for Au+Au collisions
at $\sqrt{s_{_{NN}}}$ = 200 GeV
defined as fraction of events
for which the distance between the reconstructed
and true vertices does not exceed 0.5~cm, 0.1~cm and 3~cm for the
three methods: {\it spectrometer 2D+1D} (a), {\it vertex detector} (b) 
and {\it octagon} (c). 
The additional events with incorrectly reconstructed vertex
are represented by the shaded area. The maximal allowed  distances 
between the reconstructed
and true vertices are adjusted to the resolution of the method.}
\end{figure}

The PHOBOS vertex detector is optimized for measurements of collisions occurring close
to the nominal interaction point ($Z_v$=0), but the acceptance of some of the 
reconstruction methods extends over much larger $Z_v$ range (Fig.~5).
The {\it octagon} method is the most efficient (about 95\% for the inclusive 
sample of Au+Au collisions)
and for all events, in which the vertex can also
be reconstructed using other, less efficient methods, it returns the correct position.
All methods are optimized to reconstruct the vertex
even for events with small multiplicities, thus for a fraction of events they 
fail to find the correct vertex position. In the case of the {\it spectrometer 2D+1D} 
method that results in a larger than typical error, while  
the {\it vertex detector} method quite frequently finds a vertex close to $z$=0
in the events, where the true~$Z_v$ is outside the
acceptance of this detector. Fortunately, for the later case the {\it octagon}
method gives a different vertex position and, thus, such events can be 
eliminated from the analysis.

\begin{table}[htb]
\begin{tabular}{|l|c|c|c|c|c|c|c|}
\hline
 &  \multicolumn{3}{c|}{15\% most central }  &
              \multicolumn{4}{c|}{30\% most peripheral }  \\
Method  &  \multicolumn{3}{c|}{ events}  &
              \multicolumn{4}{c|}{events}  \\
  \cline{2-8}
        &  $\sigma(X_v)$  &  $\sigma(Y_v)$  &  $\sigma(Z_v)$  
        &  $\sigma(X_v)$  &  $\sigma(Y_v)$  &  $\sigma(Z_v)$  &  eff. \\
\hline
{\it Spec. 3D}    & 0.015 & 0.022 & 0.020 &  0.350 & 0.100 & 0.300 & ~4\% \\
{\it Spec. 2D+1D} & 0.025 & 0.022 & 0.030 &  0.150 & 0.150 & 0.250 & ~7\% \\
{\it Vertex det.} & -     & 0.015 & 0.006 &    -   & 0.030 & 0.023 & 28\% \\
{\it Octagon}       & -     &   -   & 0.500 &   -   &   -   & 1.100 & 85\% \\
\hline
\end{tabular}
\vspace{0.1cm}
\caption[]{
\label{Tab1}
Vertex position accuracy ([cm]) and efficiency for Au+Au collisions
at $\sqrt{s_{_{NN}}}$ = 200 GeV in the vertex range limited to $|Z_v|<$10~cm.
To the sample of the most central events 
belong these with the largest values of the number of charged primary particles
hitting the octagonal multiplicity detector, to the peripheral sample
these with the smallest multiplicities.
For the most central events the  efficiency of all methods is 100\% (not shown).
In order to reduce the contribution of the tails of large errors the quoted
resolution is taken from the fit of a Gaussian rather than a simple RMS of the
$Z_{method}-Z_{v}$ distribution. }
\end{table}

The reconstruction error depends strongly on the multiplicity of the events, 
which becomes evident if from the sample of Au+Au collisions the most central 
and the most peripheral events are selected and separately analyzed (Table 1). 
In the first subsample all methods are 100\% efficient and 
the {\it vertex detector} method 
is the most accurate in determining $Y_v$ and $Z_v$. For peripheral
events only the {\it octagon} method has a large efficiency (85\%), whereas
both {\it spectrometer} methods are inefficient (4-7\%). The error of the vertex position
increases for all methods, about a factor of 2 for the {\it octagon} and more than 
a factor of 10 for the {\it spectrometer 3D} method. The {\it vertex detector} method 
is able to reconstruct many more events than the {\it spectrometer} methods and
still gives the most accurate vertex position. 

In order to compare reconstruction results from Monte Carlo simulations and real data,
we also calculate the values of $\sigma(Z_{method}-Z_{vertex\_method})$ 
for {\it spectrometer 3D}, {\it spectrometer 2D+1D} and {\it octagon} methods using 
events, in which all methods have found the vertex (about 50\% of
all Au+Au events). In the case of the {\it spectrometer} methods
the errors agree very well. The error for the {\it octagon} method is about 20\% smaller
in the real data than in the simulations.
This can be explained by an inaccurate description of the energy loss distribution
or by a difference in the contribution of background particles
in the GEANT simulations. Especially the slightly longer tails
of the energy loss distributions in the simulations can lead to
the observed larger error
in the calculations of the {\it octagon} vertex position.

\begin{figure}[bt]
\includegraphics[width=7.5cm]{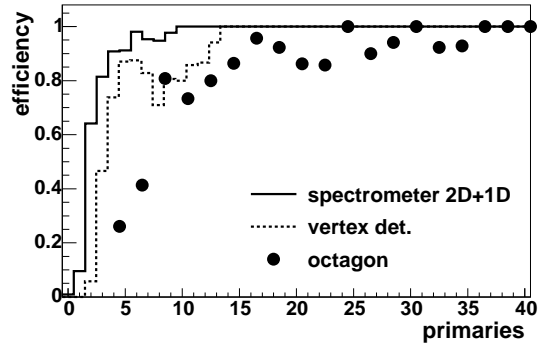}
\caption{ \label{Fig6}
The vertex reconstruction efficiency as a function of the 
number of primary charged particles present in the subdetector used
by {\it spectrometer 2D+1D}, {\it vertex detector} and {\it octagon} methods.}
\end{figure}

The efficiencies of the vertex reconstruction methods used in the PHOBOS experiment
reflect the acceptance of the subdetectors used by the reconstruction algorithm, 
mentioned in the Introduction.
To compare the ``reconstruction power'' of the 
methods using one, two or several layers of silicon we calculate
their efficiencies as the function of the number of primary
particles which point to the vertex and are registered in the appropriate 
subdetector (Fig.~6).
The high quality of the information from reconstructed tracks causes that 
{\it spectrometer 2D+1D} method can find the vertex in more than 60\% of events, 
in which only two primary particles traverse the spectrometer, and
from 4 such tracks it is over 90\% efficient. 
Interestingly, sometimes the vertex is also found in the events with only one
primary particle, but then at least one non-primary particle, for example from
a decay very close to the vertex, helped in the reconstruction.
For the {\it vertex detector} method,
as enforced by the reconstruction requirements, 
the minimal number of primary particles is 3 and 100\% efficiency is
reached for 15 such tracks. The {\it octagon} method needs at least 5 primary
particles and is more than 80\% efficient if 10 primary particles are registered
in the octagonal multiplicity detector.

\begin{table}[tb]
\begin{tabular}{|l|c|c|c|c|c|}
\hline
Colliding & $\sqrt{s_{_{NN}}}$ &  $<N_{ch}>$  &  
          \multicolumn{3}{|c|}{Efficiency} \\
\cline{4-6}
System &  [GeV]  &     & {\it spec. 2D+1D}  &  {\it ~vertex~}   &  {\it ~octagon~}  \\
\hline
Au+Au & 200      & 947 &  64.2\%       &  76.8\%  &   94.4\%  \\  
Au+Au & ~19      & 322 &  57.5\%       &  72.7\%  &   89.3\%  \\
Cu+Cu & 200      & 296 &  49.0\%       &  64.9\%  &   88.9\%  \\
d+Au  & 200      & ~66 &  12.4\%       &  39.9\%  &   86.4\%  \\
p+p   & 200      & ~16 &  ~1.4\%       &  ~6.4\%  &   53.4\%  \\
\hline
\end{tabular}
\vspace{0.1cm}
\caption{
\label{Tab2}
The efficiency of the vertex reconstruction algorithms for different collision types.
The vertex range was limited to $|Z_v|<$10~cm. The mean number of charged 
primary particles with hits in any subdetector is given as an approximation 
of the number of produced charged particles in the analyzed samples.}
\end{table}

Finally, it is worth to mention the overall efficiency of the vertex
reconstruction methods for different types of interactions measured by
the PHOBOS experiment (Table 2). For collisions of heavier nuclei (Au+Au or Cu+Cu)
the efficiencies of the precise {\it spectrometer} and {\it vertex detector} methods
are close to 50\% or higher. In the case that deuteron is one of the projectiles,
the efficiency drops faster for the {\it spectrometer} method, and becomes very small 
for both methods in the case of p+p collisions. The {\it octagon} method has
an efficiency of more than 86\% for nucleus+nucleus and more than 50\% for p+p 
interactions.

\section{Summary}
The PHOBOS experiment utilizes several vertex reconstruction algorithms, which
use complementary information from different subdetectors. According to the
design goals, a very precise vertex resolution is achieved in high 
multiplicity heavy ion
collisions, but the limited acceptance of the spectrometer and vertex detector
leads to low efficiency in peripheral Au+Au collisions, d+Au and p+p interactions.
To overcome this problem a new, very efficient method of vertex reconstruction
using hits from a single layer octagonal multiplicity detector
is introduced. Even for low multiplicity events it efficiently finds 
the vertex with an error of about 1~cm. This method can either be used 
to verify doubtful results of the other algorithms or directly in studies,
for which such precision is sufficient. Such procedure for testing 
the vertices from different methods is implemented in the PHOBOS 
experiment software. It rejects wrong reconstruction results and calculates the
most accurate vertex position.  

{
\vspace{0.4cm}
\noindent{\bf Acknowledgments} 
\vspace{0.1cm}\\
This work was partially supported by U.S. DOE grants
DE-AC02-98CH10886,
DE-FG02-93ER40802, 
DE-FC02-94ER40818,  
DE-FG02-94ER40865, 
DE-FG02-99ER41099, and
W-31-109-ENG-38, by U.S. 
NSF grants 9603486, 
0072204,            
and 0245011,        
by Polish KBN grant 1-P03B-062-27(2004-2007),
by NSC of Taiwan Contract NSC 89-2112-M-008-024, and
by Hungarian OTKA grant (F 049823).
}
%
%

\end{document}